\documentclass[aps,prd,superscriptaddress,preprint,floatfix,nofootinbib]{revtex4}
\usepackage{graphicx}  
\usepackage{dcolumn}   
\usepackage{bm}        
\usepackage{amssymb}   
\usepackage{amsmath}
\usepackage{epsfig}
\usepackage[hidelinks]{hyperref}
\usepackage{rotating}
\usepackage{slashed}
\usepackage[usenames, dvipsnames]{color}
\newcommand{\be}{\begin{equation}}
\newcommand{\ee}{\end{equation}}
\newcommand{\nn}{\nonumber}
\numberwithin{equation}{section}

\makeatletter
\renewcommand{\p@subsection}{}
\renewcommand{\p@subsubsection}{}
\def\l@subsubsection#1#2{}
\makeatother
\begin{document}
	
	
	\title{Parity-Violating Gravity and GW170817 in Non-Riemannian Cosmology}
	
	\author{Aindri\'u Conroy}
	\affiliation{Centre for Astrophysics and Relativity,
		School of Mathematical Sciences,
		Dublin City University,
		Glasnevin,
		Dublin 9, Rep. of Ireland.
	}
	\affiliation{Department of Physics, Lancaster  University, Lancaster, LA1 4YB, UK}
	\author{Tomi Koivisto}
	\affiliation{Laboratory of Theoretical Physics, Institute of Physics, University of Tartu, W. Ostwaldi 1, 50411 Tartu, Estonia}
	\affiliation{National Institute of Chemical Physics and Biophysics, R\"avala pst. 10, 10143 Tallinn, Estonia}
	\date{\today}
	
	\begin{abstract}
		The cosmological propagation of tensor perturbations is studied in the context of parity-violating extensions of the symmetric teleparallel equivalent of General Relativity theory.
This non-Riemannian formulation allows for a wider variety of consistent extensions than the metric formulation of gravity theory. It is found that while many of the 
possible quadratic terms do not influence the propagation of the gravitational waves, a generic modification predicts a signature that distinguishes the left- and right-handed 
circular polarizations. The parameters of such modifications can be constrained stringently because the propagation speed of the gravitational waves is scale-dependent    
and differs from the speed of light.
	\end{abstract}
	
	\maketitle
	\tableofcontents
	\section{Introduction}
		A new era of observational physics is flourishing with the discovery of gravitational waves (GW) at the LIGO detector \cite{TheLIGOScientific:2017qsa}. These waves are becoming an increasingly important probe of the universe and its dynamics. One of the interesting applications of the GW170817/GRB170817A and further such data is its use to constrain the theory of gravity. In particular, the propagation speed of GW has been measured quite 
	precisely by comparing the arrival times between the gravitational signal from the merger of neutron stars and a short gamma-ray burst of high-energy photons. Therefore, modified gravity models which predict appreciable 
	differences between the propagation speed of GW and the speed of light can now be ruled out. As a result, we can now exclude a large class of scalar-tensor models as a means of explaining the acceleration of cosmic expansion \cite{Ezquiaga:2017ekz,Amendola:2017orw}. A plethora of other types of modified gravity models have been proposed \cite{Heisenberg:2018vsk} and the emerging new field of multi-messenger GW astronomy offers a variety of experimental probes, in addition to propagation speed, which can be used to test such models \cite{Nishizawa:2017nef,Ezquiaga:2018btd}. 
	    
	In this paper we are interested in constraining violations of parity in the gravitational sector with the physics of GW. Parity violations feature often in considerations of quantum gravity \cite{Mavromatos:2004sz,Freidel:2005sn} and 
	unification of gravity with the particle interactions \cite{Alexander:2012ge,Krasnov:2017epi}. Regarding observational constraints on parity-violating gravity, the case of a primordial spectrum of GW \cite{Creminelli:2014wna} produced at inflation would be difficult to detect from the angular spectrum \cite{Gerbino:2016mqb},
	but in principle there are models whose predictions could be tested using correlations of spectra and higher order spectra  \cite{Zhu:2013fja,Masui:2017fzw,Bartolo:2017szm}. The direct detections of GW offer new possibilities.  
	For example, in an axiverse-motivated scenario, if a parametric resonance occurs due to coherent axion oscillations, the circular polarization of GW induced by the Chern-Simons coupling could become detectable \cite{Yoshida:2017cjl}. 
	Recently, the propagation of GW was considered in a generic parity-violating metric gravity theory \cite{Nishizawa:2018srh}. The conclusion was that unless the theory is reduced to the special case of the Chern-Simons coupling, the propagation speed of GW is modified and therefore such parity-violating corrections to the Einstein-Hilbert term are stringently constrained.  
	
	   Due to the higher-order property of the Riemann curvature,  the Einstein-Hilbert action hides second derivatives and its generalisations are severely restricted. Combining this Riemannian formulation with the metric teleparallel \cite{Aldrovandi:2013wha} and the symmetric teleparallel \cite{Nester:1998mp} equivalents of General Relativity forms the Geometric Trinity \cite{Heisenberg:2018vsk}. The latter two formulations  appear to provide a more flexible framework for generalisations, since their action principles feature only first derivatives.   
	  Taking the coincident General Relativity \cite{BeltranJimenez:2017tkd} as the starting point, we consider parity violations in terms of non-metricity. As the non-metricity tensor is first order in derivatives, there is in principle 
	  an infinite number of terms one could consider without resorting to higher derivatives. However, in this paper we shall explore only the effect of generic parity-violating quadratic corrections to the propagation of GW. 
	GW have been considered at many occasions in extended symmetric teleparallel gravity theories \cite{Adak:2008gd,BeltranJimenez:2017tkd,Conroy:2017yln,BeltranJimenez:2018vdo,Hohmann:2018xnb,Hohmann:2018wxu,Soudi:2018dhv,Jimenez:2019ovq}, but the possibility of parity violations has not been taken into account. Symmetric teleparallel gravity has been recently discussed in the context of dark energy at cosmological scales \cite{Jarv:2018bgs,Jimenez:2019ovq,Lu:2019hra,Lazkoz:2019sjl} and dark matter at galactic scales \cite{Milgrom:2019rtd}. 
	
	We begin the paper by reviewing some basic equations of symmetric teleparallel cosmology in Section \ref{sec:prelim}, and then in the following two sections, we give a complete analysis of all possible parity-violating terms that are quadratic in non-metricity, deriving finally the constraints on all those terms in the light of the GW170817/GRB170817A in Section
	\ref{propagation}. We then discuss our conclusions in Section \ref{sec:conc}. 
	
	\section{Mathematical Preliminaries}
	\label{sec:prelim}
		\subsection{Perturbed non-metricity}
		\label{sec:pert}
	The non-metricity tensor and its contractions are defined by
	\begin{equation}
	\label{Qdefs}
	Q_{abc}\equiv\nabla_a g_{bc}\,,\qquad Q_a\equiv g^{bc}Q_{abc}\,,\qquad \tilde Q_c \equiv g^{ab}Q_{abc}\,,
	\end{equation}
	where our convention is that Latin indices towards the start of the alphabet $a,b,c,d,e,f$ include both spatial and temporal components, while indices towards the middle $i,j,k,l,m$ denote spatial components only. The covariant derivative $\nabla$ is with respect to a generic connection $\Gamma$. Taking the variation of the non-metricity tensor $Q_{abc}$, we find
	\begin{equation}
	\delta Q_{abc}	= \nabla_{a}h_{bc}-\gamma{}_{cab}-\gamma{}_{bac}\,,
	\end{equation}
	where we have defined  $\delta \Gamma^c{ }_{ab}\equiv \gamma^c{}_{ab}$ and are perturbing according to $g_{ab} \rightarrow \bar{g}_{ab} + \epsilon h_{ab}$, $\Gamma^{a}{}_{bc}\rightarrow\bar{\Gamma}^{a}{}_{bc}+\epsilon\gamma^{a}{}_{bc}$. Here and throughout, `barred' quantities indicate their evaluation on the background.	
	It is helpful also to define the perturbed non-metricity tensor with one index up, like so $q_{ab}{}^{c}\equiv\delta Q_{ab}{}^{c}$ and as such we write
	\begin{align}
	\label{qabc0}
	q_{abc}&=-h_{cd}\bar{Q}_{ab}{}^{d}+{\nabla}_{a}h_{bc}-\gamma{}_{cab}-\gamma{}_{bac}=g_{cf}\delta Q_{ab}{}^{f}\,,
	\nn\\
	q_{a}&=-h^{b}{}_{d}\bar{Q}_{ab}{}^{d}+\bar{g}^{bc}{\nabla}_{a}h_{bc}-2\gamma^{b}{}_{ab}=\delta Q_{a}\,,
	\nn\\
	\tilde{q}_{c}&=-h^{ab}\bar{Q}{}_{abc}-h_{cd}\tilde{\bar{Q}}{}^{d}+{\nabla}^{b}h_{bc}-2\gamma^{b}{}_{cb}=\delta\tilde{Q}_{c}\,.
	\end{align}
 We note that while the contraction $\bar g^{bc}q_{abc}=q_a$ holds in the perturbed case as in \eqref{Qdefs}, the contraction $\bar g^{ab}q_{abc} \neq \tilde{q}_a$ in general does not. 
 
 In this work we will specialise to the symmetric teleparallel geometry \cite{Nester:1998mp}. Thus, the connection $\Gamma$ is considered to be devoid of both curvature and torsion. This allows us to choose
 the coincident gauge \cite{BeltranJimenez:2017tkd} wherein we can simply write partial derivatives in place of the covariant operators above.
 Around a Friedmann-Robertson-Walker (FRW) background,
	\begin{equation}
	\label{FRW}
	ds^{2}=-dt^{2}+a^{2}(t)\delta_{ij}dx^{i}dx^{j}\,,
	\end{equation} 
	which is our focus here, the only non-vanishing components of the non-metricity tensor in the coincident gauge are given by
	\begin{equation}
	\label{FRWcpts}
	\bar Q_{0ij}=2H\bar g_{ij}=-\bar Q^{0}{}_{ij}\,,\qquad \bar Q_{0}=6H=-\bar Q^{0}\,.
	\end{equation}
	As such, the first two terms of $\tilde q_c$ vanish and the contraction does in fact hold, $\bar g^{ab}q_{abc} = \tilde{q}_a$. 
		Taking into account the background (\ref{FRW}) as well as the relation
	\begin{equation}
	\label{delgammau}
	\delta \Gamma^a{}_{bc}
	=\epsilon\partial_b\partial_c u^a\,,
	\end{equation}
derived in Appendix \ref{sec:Lie} from the definition of the Lie derivative, we find \eqref{qabc0} to reduce to
	\begin{align}
	\label{qabc1}
	q_{abc}	&=-2H\delta_{a}^{0}h_{bc}+\partial_{a}h_{bc}-2g_{(bd}\partial_{a}\partial_{c)}u^{d}\,,
	\nn\\q_{a}	&=-2H\bar{g}^{bc}\delta_{a}^{0}h_{bc}+\bar{g}^{bc}\partial_{a}h_{bc}-2\partial_{a}\partial_{d}u^{d}\,,
	\nn\\\tilde{q}_{c}	&=\partial^{f}h_{fc}-2\bar{g}^{af}\bar{g}_{(fe}\partial_{a}\partial_{c)}u^{e}\,.
	\end{align}
	One can compare these perturbations with those found in \cite{Conroy:2017yln} around Minkowski space at the limit $H\rightarrow 0$. 
	In the following, we shall specialise to spatial perturbations around the background (\ref{FRW}).

	\subsection{Perturbed Continuity Equation}
	\label{sec:cont}
	
	When considering a modified teleparallel gravity theory, it is crucial to take into account the equation of motion for the connection, which can be shown to be equivalent to the covariant conservation of the matter
	energy momentum \cite{BeltranJimenez:2018vdo} (and in metric teleparallelism, to the equation of motion of the antisymmetric components). Thus, only if we can show that the matter conservation is retained in a modified theory, can we use the metric field equations to determine the dynamics
	of the theory without additional constraints from the connection equation of motion. Therefore it is important to carefully take into account the continuity equation when considering cosmological modifications of gravity.
	
		In order to study the propagation of GW, we need only perturb the spatial portions of the metric like so
	\begin{equation}
	\label{gpert}
	g_{ij}\rightarrow\bar{g}_{ij}+\epsilon h_{ij}\,,
	\end{equation}
	where $\bar g$ is the metric tensor for the FRW metric \eqref{FRW}
	and $\epsilon$ is an infinitesimal parameter. In terms of the perturbation of the Levi-Civita connection
		\begin{equation}
	\label{pertgamma}
	\hat\gamma^a{}_{bc}=-\frac{1}{2}{h}^{ad}(\partial_{b}\bar{g}_{cd}+\partial_{c}\bar{g}_{bc}-\partial_{d}\bar{g}_{bc}) + \frac{1}{2}\bar{g}^{ad}(\partial_{b}h_{cd}+\partial_{c}h_{bc}-\partial_{d}h_{bc})\,,
	\end{equation}
	we must perturb both spatial and temporal indices but remain mindful that only spatial indices may enter the perturbed metric tensor $h_{ab}$. With this in mind, we may read off the non-vanishing components of \eqref{pertgamma} like so
	\begin{equation}
	\hat\gamma^{0}{}_{ij}=\frac{1}{2}\partial_{t}h_{ij}\,,\qquad\hat\gamma^{k}{}_{ij}=\frac{1}{2}\bar{g}^{kl}(\partial_{i}h_{lj}+\partial_{j}h_{il}-\partial_{l}h_{ij})\,,\qquad\hat\gamma^{i}{}_{0j}=\frac{1}{2}\bar{g}^{il}\partial_{t}h_{lj}\,.
	\end{equation}
		We use these relations in order	to express the perturbed continuity equations. Perturbing the divergence $\hat \nabla_a T^a { }_b=0$ to linear order around an FRW background we obtain
	\begin{equation}
	\hat{\nabla}_{a}T^{a}{}_{b}	=-\delta_{b}^{0}\left(\dot{\bar{\rho}}+3H(\bar{\rho}+\bar{p})\right)+\epsilon\left(\delta_{b}^{j}\partial_{i}-\delta_{b}^{0}H\delta_{i}^{j}\right)\tau^{i}{}_{j}+{\cal O}(\epsilon^{2})=0\,,
	\end{equation}
	where $\hat\nabla$ is the covariant derivative for the Levi-Civita connection and the matter content of the Universe, evaluated on the background, is given by $\bar{T}^{0}{}_{0}=-\bar{\rho}, \delta_{j}^{i}\bar{T}_{i}^{j}=3\bar{p}$, with $\rho$ being the energy density and $p$ the pressure of matter. Furthermore, the perturbed stress-energy tensor obeys $\tau^0{}_0=\tau^i{}_0=0$, which results from the perturbing of only the spatial indices of the metric tensor. We then find
	\begin{equation}
	-\left(\dot{\bar{\rho}}+3H(\bar{\rho}+\bar{p})\right)-\epsilon H\delta_{i}^{j}\tau^{i}{}_{j}+{\cal O}(\epsilon^{2})=0\quad \iff\quad\epsilon\delta_{k}^{j}\partial_{i}\tau^{i}{}_{j}+{\cal O}(\epsilon^{2})=0\,.
	\end{equation}
	As such, the continuity equation is satisfied to linear order for a spatially perturbed stress-energy tensor that is transverse and traceless, i.e.
	\begin{equation}
	\label{pertcontcond}
	\partial_i\tau^i{}_j=0\,,\qquad \delta^j_i \tau^i{}_j=0\,.
	\end{equation}
	We will return to these relations later to show the parity-violating field equations we shall derive below for extended symmetric teleparallel gravity theories are indeed conserved in a minimally coupled system.

         \subsection{The non-metricity equivalent of General Relativity}
         \label{sec:qgr}
         
         Having defined non-metricity and its perturbations  in Section \ref{sec:pert}, let us briefly review the construction of the basis of such a theory. We shall denote the
         gravitational coupling constant by $\kappa=8\pi G_N$, $G_N$ being the Newton's constant. 
         An action 
         \be \label{qgr}
         S_{QGR}=-\frac{1}{2\kappa}\int d^4 x \sqrt{-g}L_{QGR}\,,
         \ee
         that is equivalent to General Relativity, can be specified as \cite{BeltranJimenez:2017tkd,Conroy:2017yln}
         \be \label{qgrscalar}
         L_{GQR} = -\frac{1}{4}Q_{abc}Q^{abc} + \frac{1}{2}Q_{abc}Q^{bac} + \frac{1}{4}Q_a Q^a - \frac{1}{2}Q_a\tilde{Q}^a\,.
         \ee 
         If the curvature scalar of the connection is denoted by $R$, one may show that 
         \be
         R = \hat{R} + L_{GQR} + \hat{\nabla}_a\left( Q^a - \tilde{Q}^a\right)\,,
         \ee 
         and thus (\ref{qgr}) is related to the Einstein-Hilbert action by a total derivative. In the following, we will then consider quadratic, parity-violating corrections to (\ref{qgr}), and in particular determine their impact to the propagation of GW.

	\section{Parity Violation in Non-Riemannian Cosmology}
	We seek to investigate parity-violating gravity that is quadratic in non-metricity in the symmetric teleparallel geometry. Without coupling to a scalar field, the sole surviving terms are given by
	\begin{equation} \label{pvterms}
	S_{PV}=\frac{1}{2\kappa}\int d^{4}x\sqrt{-g}\varepsilon^{abcd}\left(Q_{abe}\alpha Q_{cd}{}^{e}+\nabla_{a}Q_{efb}\beta\nabla_{c}Q^{ef}{}_{d}+\nabla_{a}\tilde{Q}_{b}\gamma\nabla_{c}\tilde{Q}_{d}\right)\,.
	\end{equation}
	We see here that the parity-violating gravity in STG contains terms that are both second-order and fourth-order in derivatives. Further higher-derivative extensions could be attained by promoting the constants $\alpha, \beta, \gamma$ to functions of covariant operators, i.e. $\alpha=\alpha(\Box), \beta=\beta(\Box), \gamma=\gamma(\Box)$ (see infinite derivative theories such as e.g. \cite{Biswas:2005qr, Biswas:2011ar,Biswas:2013cha,Conroy:2017uds}) but this lies outside the present study. In addition to the restricting to quadratic models, we shall also
	only focus on models that feature (at most) second time-derivatives, in order to potentially exclude ghosts. 
	
When coupling to a scalar field, there are three distinct classes of Lagrangian to investigate, which are quadratic in non-metricity:
\begin{equation}
L_a\sim \varepsilon^{abcd}\phi\phi QQ,\quad L_A=\varepsilon^{abcd}\phi Q\nabla Q,\quad L_B=\varepsilon^{abcd}\phi \phi \nabla Q \nabla Q.
\end{equation}
$L_a$ denotes a second-derivative theory, whereas the capitals $A,B$ denote a higher order theory.
	Here, $\varepsilon$ is the Levi-Civita symbol and $Q$ represents all possible forms of the non-metricity tensor $Q_{abc}=\nabla_a g_{bc}$, where we have suppressed the indices in order to give a schematic description of the methodology. 
	
	\subsection{The generic second order Lagrangian}
	
	Let us begin by concentrating on the first class of Lagrangians, which are quadratic in non-metricity (like the others), coupled quadratically to a scalar field and second-order in derivatives (and not only time derivatives). 
	Within this second-derivative class of Lagrangians $L_a$, we analyse all possible permutations including contractions with both the metric tensor $g_{ab}$ and the Levi-Civita symbol $\varepsilon^{abcd}$. From the symmetries contained within the Lagrangian along with the antisymmetric properties of the Levi-Civita symbol, we can establish the following rules to aid us in the process: All terms with
	\begin{enumerate}
		\item ${abcd}$ appearing twice in the scalar field $\phi$, or
		\item ${abcd}$ appearing twice in the last two indices of the non-metricity tensor $Q_{abc}$,
	\end{enumerate}
	will vanish. Consequently, we then obtain 7 non-vanishing Lagrangians:
	\begin{equation*}
	\begin{aligned}[b]
	\label{Lagrangians17}
	L_{a1}	&=\varepsilon^{abcd}\phi^{e}\phi^{f}Q_{abe}Q_{cdf}
	\nn\\	
	L_{a2}	&=\varepsilon^{abcd}\phi_{c}\phi^{f}Q_{abe}Q^{e}{}_{df}
	\nn\\
	L_{a3}	&=\varepsilon^{abcd}\phi_{c}\phi^{f}Q_{abe}Q_{fd}{}^{e}
	\nn\\L_{a4}&=\varepsilon^{abcd}\phi_{c}\phi^{f}Q_{abe}Q_{df}{}^{e}
	\end{aligned}
	\qquad
	\begin{aligned}[b]
		\nn\\L_{a5}	&=\varepsilon^{abcd}\phi_{c}\phi^{e}Q_{abe}Q_{d}
	\nn\\L_{a6}&=\varepsilon^{abcd}\phi_{c}\phi^{e}Q_{abe}\tilde{Q}_{d}
	\nn\\L_{a7}	&=\varepsilon^{abcd}\phi_{f}\phi^{f}Q_{abe}Q_{cd}{}^{e}\,.
	\end{aligned}
	\end{equation*}
	We may then write the action for parity-violating terms coupled to a scalar field like so
	\begin{equation}
	S_{PV}^{(2)}=\frac{1}{2\kappa}\int d^{4}x\sqrt{-g}\alpha_{i}L_{ai}\,,
	\end{equation}
	where the $^{(2)}$ indicates that, at this stage, we are considering only terms of second-order in derivatives and $\alpha_{i}(\phi,\phi_{a}\phi^{a})$ is an arbitrary function of field $\phi$ and its kinetic term.

	\subsubsection{The unique and non-vanishings Lagrangians for GW}
	
	We must now determine which of the 7 Lagrangians above are unique and non-vanishing once perturbations of the spatial indices of the metric tensor have been performed, see Section \ref{sec:pert}. We can reduce the number of non-vanishing Lagrangians greatly by noting that around an FRW metric, both $Q_{abc}$ and its (spatial) perturbation $q_{abc}$ cannot have $0$ in the last two indices, which can be easily seen from \eqref{qabc1}. 
	This combined with the fact that as $\phi_a =\partial_a \phi(t)$, the index of the scalar field must be temporal (i.e. $\phi_a=\delta^0_a\dot{\phi}$), we find that only two Lagrangians remain,
	\begin{equation}
	\label{L2deriv}
	L_{a3}=\varepsilon^{abcd}\phi_{c}\phi^{f}Q_{abe}Q_{fd}{}^{e},\qquad L_{a7}=\varepsilon^{abcd}\phi_{f}\phi^{f}Q_{abe}Q_{cd}{}^{e}\,.
	\end{equation}	
	Beginning with $L_{a3}$, we perturb according to $Q_{aj}{}^{k}\rightarrow\bar{Q}_{aj}{}^{k}+\epsilon q_{aj}{}^{k}$ which follows from \eqref{gpert}, like so 
	\begin{align}
	L_{a3}&=\varepsilon^{abcd}\phi_{c}\phi^{f}\left[\bar Q_{aeb}\bar Q_{fd}{}^{e}+\epsilon\left(q_{aeb}\bar Q_{fd}{}^{e}+\bar Q_{aeb}q_{fd}{}^{e}\right)+\epsilon^{2}q_{aeb}q_{fd}{}^{e}\right]
	\nn\\&=\phi_{0}^{2}\left(\epsilon\varepsilon^{ijtk}q_{ilj}\bar Q_{0k}{}^{l}+\epsilon^{2}\varepsilon^{ijtk}q_{ilj}q_{0k}{}^{l}\right)
	\nn\\&=\phi_{0}^{2}\left(\epsilon\varepsilon^{tijk}2Hq_{ikj}+\epsilon^{2}\varepsilon^{tijk}q_{ilj}q_{0k}{}^{l}\right)\,,
	\end{align}
	where we have noted that $\bar Q_{0k}{ }^l=2H\delta_k ^l$ from \eqref{FRWcpts}. We then note that the ${\cal O}(\epsilon)$ term vanishes due to the symmetry of $q_{ikj}$ in the last two indices and the antisymmetric properties of the Levi-Civita symbol, leaving
	\begin{equation}
	L_3=\epsilon^{2}\varepsilon^{tijk}\phi_{0}^2q_{ilj}q_{0k}{}^{l}\,.
	\end{equation}
	Repeating the process for $L_{a7}$ we find that both Lagrangians are essentially the same, differing only through a constant, i.e. $L_{a3}=2L_{a7}$ and, as such, we write
	\begin{equation}
	{\cal L}_{PV1}^{(2)}\supset\varepsilon^{ijk}\dot\phi^2q_{ijl}q_{0k}{}^{l}\,.
	\end{equation} 
		For the propagation of GW, we need only consider the metric fluctuations and, as such, the relevant terms from \eqref{qabc1} are simply
	\begin{equation}
	q_{ijl}=\partial_{i}h_{jl}\,,\qquad q_{0k}{}^{l}=-2Hh_{k}{}^{l}+\bar{g}^{jl}\partial_{t}h_{kj}\,. 
	\end{equation}
	Substitution into the surviving Lagrangian gives
	\begin{equation} \label{aterm1}
	{\cal L}_{PV1}^{(2)}\supset\varepsilon^{ijk}\dot\phi^2\left(-2Hh_{k}{}^{l}+\dot{h}_{km}\bar{g}^{ml}\right)\partial_{i}h_{jl}\,.
	\end{equation}
	Note that this corresponds to the first term in (\ref{pvterms}), when the parameter $\alpha$ is replaced by the kinetic term of a scalar field. In general one could consider the $\alpha$ to be a function of the background 
	(through an additional scalar field or otherwise). The form of the Lagrangian (\ref{aterm1}) captures the generic leading order effect of parity-violating non-metricity on spatial perturbations of the metric, and in the following we shall employ the parameterisation
	\begin{equation} \label{aterm}
	{\cal L}_{PV1}^{(2)} = \varepsilon^{ijk}\frac{\alpha(t)}{a}\left(-2Hh_{k}{}^{l}+\dot{h}_{km}\bar{g}^{ml}\right)\partial_{i}h_{jl}\,,
	\end{equation}
	where $\alpha$ is considered to be a dimensionless, time-dependent function.

	\subsection{Propagation of GW}
	\label{sec:GW2nd}
	
	From here on we will restrict to tensor fluctuations, i.e. the perturbations around FRW that are transverse and traceless. We can therefore set $\partial_i h^i{}_j=0$ and $h^i{}_i=0$ in order to satisfy the continuity equation, see Section \ref{sec:cont}.
	
	Now, as $\bar{g}^{jl}$ is dependent only on $t$, we have $\bar{g}^{jl}\partial_{i}h_{jl}=\partial_{i}(\bar{g}^{jl}h_{jl})=\partial_{i}h$, and due to the transverse-traceless restriction, the second term in (\ref{aterm}) vanishes leaving only the first.
	Thus, we may write
	\begin{equation}
	\label{S2action}
	S^{(2)}=\frac{1}{2\kappa}\int dt\;d^{3}x\;a^{3}\left({\cal L}_{QGR}^{(2)}+{\cal L}_{PV}^{(2)}\right)\,,
	\end{equation}
	where
	\begin{equation}
	{\cal L}_{PV}^{(2)}=\frac{H}{a}\alpha(t)\varepsilon^{ijk}h_{k}{}^{l}\partial_{i}h_{jl}\,,\quad\mbox{and}\quad{\cal L}_{QGR}^{(2)}=\frac{1}{4}\left(\dot{h}^{ij}\dot{h}_{ij}-\partial^{k}h_{ij}\partial_{k}h^{ij}\right)\,,
	\end{equation}
	for some dimensionless function of time $\alpha(t)$. 
	${\cal L}^{(2)}_{QGR}$ is derived from the non-metric equivalent to the Einstein-Hilbert action we
	recall from Section \ref{sec:qgr},
	and is precisely equivalent to GR, see \cite{Jimenez:2019ovq} for generalisation of the result for a nonlinearised action\footnote{In such models, the propagation speed of GW is not modified, but the expansion friction term is modified (to be later parameterised by $\nu$). Such modification could also be constrained by the future data on GW, as one can deduce from a recent study of models that have the same evolution of the background expansion and of the GW \cite{Nunes:2019bjq}.}, $L_{QGR}\rightarrow f(L_{QGR})$.  

	\subsubsection*{Field equations: non-metric equivalent to General Relativity}
	We shall now derive the field equations from the above combined action, starting with the well-known part as a warm-up. 	To begin, rewrite ${\cal L}_{QGR}$ in Fourier space like so
	\begin{equation}
	S_{QGR}^{(2)}	=\frac{1}{2\kappa}\frac{1}{4}\int d^{3}x\;dt\;a^{3}\left(\partial_{t}h^{ij}\partial_{t}h_{ij}-a^{-2}\delta^{kl}\partial_{l}h_{ij}\partial_{k}h^{ij}\right)\,,
	\end{equation}
	and vary w.r.t. $h^{ij}$,
	\begin{equation}
	\delta S_{QGR}^{(2)}	=\frac{1}{2\kappa}\frac{1}{4}\int d^{3}x\;dt\;\left(-2\partial_{t}(a^{3}\partial_{t}h_{ij})+2a\delta^{kl}\partial_{k}\partial_{l}h_{ij}\right)\delta h^{ij}\,.
	\end{equation}
	Transforming into conformal coordinate using $d\eta=a^{-1}dt$, where $\partial_t f=a^{-1}f^\prime$, ${\cal H}=a^\prime /a$ and $^\prime$ denotes the derivative w.r.t. conformal time, gives
	\begin{equation}
	\delta S_{QGR}^{(2)}	=-\frac{1}{2\kappa}\frac{1}{2}\int d^{3}x\;d\eta\sqrt{-g_{c}}\;a^{-2}\left(2{\cal H}h_{ij}^{\prime}+h_{ij}^{\prime\prime}-\partial^{2}h_{ij}\right)\delta h^{ij}\,,
	\end{equation}
	where $g_c$ is the determinant of the metric in the conformal coordinates. We may then read off the field equations like so
	\begin{equation}
	\tau_{ij}^{QGR}=\frac{1}{2a^{2}\kappa}\left(h_{ij}^{\prime\prime}+2{\cal H}h_{ij}^{\prime}-\partial^{2}h_{ij}\right)\,.
	\end{equation}
	This gives the wave equation $\tau_{ij}^{QGR}=0$, which is clearly transverse and traceless, as the perturbation $h_{ij}$ is.

	\subsubsection*{Field equations: parity-violating action}
	Following the same process as above for the action
	\begin{equation}
	S_{PV}^{(2)}	=\frac{1}{2\kappa}\int d^{3}x\;d\eta a^{2}{\cal H}\alpha\varepsilon_{ijk}h{}^{kl}\partial^{i}h_{l}^{j}\,,
	\end{equation}
	we find the field equations to be
	\begin{equation}
	\tau_{ij}^{PV}=-\frac{2}{a^2\kappa}{\cal H}\alpha\varepsilon_{kl(i}\partial^{k}h_{j)}^{l}\,.
	\end{equation}
	Combining with the QGR field equations, we find
	\begin{equation}
	\tau_{ij}=\frac{1}{2a^{2}\kappa}\left(h_{ij}^{\prime\prime}+2{\cal H}h_{ij}^{\prime}-\partial^{2}h_{ij}\right)-\frac{1}{a^2\kappa}{\cal H}\alpha\varepsilon_{ikl}\partial^{k}h_{j}^{l}-\frac{1}{a^2\kappa}{\cal H}\alpha\varepsilon_{jkl}\partial^{k}h_{i}^{l}.
	\end{equation}
	We briefly note that the partial divergence equation $\partial_i \tau^{ij}=0$ is satisfied due to the transverse nature of the metric perturbations, i.e. $\partial_i h^{ij}=0$, along with the antisymmetric properties of the Levi-Civita symbol. The trace also vanishes meaning the perturbed divergence equation \eqref{pertcontcond} is satisfied and the field equations are conserved.

	\subsubsection*{Polarization basis}
	We have now established that the relevant equations of motion for the action \eqref{S2action} are given by
	\begin{equation}
	\label{fieldS2}
	\left(h_{ij}^{\prime\prime}+2{\cal H}h_{ij}^{\prime}-\partial^{2}h_{ij}\right)-4{\cal H}\alpha\varepsilon_{kl(i}\partial^{k}h_{j)}^{l}=0\,.
	\end{equation}
	Following the same prescription as \cite{Nishizawa:2018srh}, we decompose the metric perturbation into the circular polarization basis defined by
	\begin{equation}
	e_{ij}^R\equiv \frac{1}{\sqrt{2}}(e_{ij}^++ie_{ij}^\times)\,,\qquad
	e_{ij}^L\equiv \frac{1}{\sqrt{2}}(e_{ij}^+-ie_{ij}^\times)\,,
	\end{equation}
	which allows us to decompose $h_{ij}$ into Fourier space like so
	\begin{equation}
	\label{hijdecomp}
	h_{ij}(\eta,\vec{x})=\frac{1}{(2\pi)^{3/2}}\sum_{A=R,L}\int d^{3}x\;h_{\vec{k}}^{A}(\eta)e_{ij}^{A}e^{i\vec{k}\cdot\vec{x}}\,.
	\end{equation}
	Noting the relations 
	\begin{equation}
ik^{l}\varepsilon_{ilk}e_{j}^{A\;k}=ik\varepsilon_{ilk}n^{l}e_{j}^{A\;k}=-k\lambda_{A}e_{ij}^{A}\,,
	\end{equation}
	where $n_i$ is a unit vector facing the direction of propagation, $A=R,L$ and $\lambda_R=1$, $\lambda_L=-1$, before substituting \eqref{hijdecomp} into \eqref{fieldS2} gives
	\begin{equation}
	\label{GWs1}
	(h_{k}^{A})^{\prime\prime}+2{\cal H}(h_{k}^{A})^{\prime}+\left(1+\frac{4{\cal H}\alpha\lambda_{A}}{ k}\right)k^{2}h_{k}^{A}	=0\,.
	\end{equation}
	We see here that the parity-violating extension to General Relativity modifies only the final term.

	\subsubsection*{Analysis}
	
	In the general formulation of the propagation of GWs, tensor perturbations follow the equation of motion \cite{Saltas:2014dha,Sawicki:2016klv,Nishizawa:2017nef}
	\begin{equation}
	\label{GWsgeneric}
	h_{ij}^{\prime\prime}+(2+\nu){\cal H}h^\prime_{ij}+\left(c_{T}^2 k^2+a^2\mu^2\right)h_{ij}=a^2 \Gamma \gamma_{ij}\,,
	\end{equation}
	where $\nu={\cal H}^{-1}(d \ln M_*^2 /dt)$ is the effective Planck mass $M_*=\sqrt{8\pi \kappa_*^{-1}}$ run rate \cite{Amendola:2017ovw}, $c_{T}$ is the propagation speed the GW and $\mu$ is the effective mass of the graviton. Comparing the above with \eqref{GWs1}, we find that only the propagation speed of the GW is modified by the new geometry, in such a way that if the GW with right-handed polarization travels faster than light, then the GW with left-handed polarization travels slower (and vice versa, depending on the sign of $\alpha$). We can consider the dimensionless parameter $\alpha$ as quantifying the relative magnitude of the parity-violating correction. As we will show in more detail in Section \ref{pgw}, the 
	current data from the event GW170817/GRB170817A does not constrain $\alpha$. In principle, future data has the potential to detect the possible signature of the parity violation. This does not require a simultaneous light signal, but only the measurement of right-handed and left-handed polarisations arriving at slightly different times.

	\section{Higher Derivative Parity-Violating}
	
	We proceed in a similar fashion to the previous section in order to determine the unique, non-vanishing, parity-violating Lagrangians in the symmetric teleparallel geometry, this time including higher derivatives. To this end, we investigate the two classes of Lagrangian given by,
	\begin{equation}
	L_A\sim \varepsilon^{abcd} \phi Q\nabla Q\,,\qquad L_B=\varepsilon^{abcd} \phi\phi\nabla Q \nabla Q\,,
	\end{equation}
	which capture the essence of all higher-derivative extensions that are quadratic in non-metricity. In principle, there are non-vanishing terms of all orders in derivatives but as discussed in Appendix \ref{sec:nohigh}, these higher-order terms are straightforward and somewhat trivial generalisations of second and third-derivative gravity. We include the class of fourth-order Lagrangians $L_B$ to verify this explicitly.
	
	 We further note that while these higher-order terms are not guaranteed to be ghost-free, it is of interest, in the spirit of effective field theory, to take them into
	account in order to determine the most generic possible effect of parity-violating terms at the leading (quadratic) order. Also, we restrict to the case of (at most) second order
	time derivatives. 
	
	\subsection{Generic higher derivative terms}
	\label{sec:genhigh}
	Due to the symmetry $\nabla_a Q_{bcd}=\nabla_bQ_{acd}$ (which follows from the generalised Ricci identity with vanishing curvature and torsion) along with the antisymmetric properties of the Levi-Civita symbol, we may add a third rule to aid us in identifying the surviving Lagrangians, i.e. terms with
		\begin{enumerate}
		\item ${abcd}$ appearing twice in the instances of the scalar field $\phi$, or
		\item ${abcd}$ appearing twice in the last two indices of the non-metricity tensor $Q$, or
		\item ${abcd}$ appearing in $\nabla$ and the first index of $Q$,
	\end{enumerate}
will vanish. Following these rules gives twelve different Lagrangians for each of the above two classes
\begin{equation}
\begin{aligned}[c]
L_{A1}&=\varepsilon^{abcd}\phi_{d}\nabla_{a}Q_{fb}{}^{e}Q_{ce}{}^{f}
\\
L_{A2}&=\varepsilon^{abcd}\phi_{d}\nabla_{a}Q_{fb}{}^{e}Q_{ec}{}^{f}
\\
L_{A3}&=\varepsilon^{abcd}\phi_{d}\nabla_{a}Q_{fb}{}^{e}Q^{f}{}_{ce}
\\
L_{A4}&=\varepsilon^{abcd}\phi_{d}\nabla_{f}Q^{fe}{}_{a}Q_{bce}
\\
L_{A5}&=\varepsilon^{abcd}\phi_{d}\nabla_{a}\tilde{Q}^{e}Q_{bce}
\\
L_{A6}&=\varepsilon^{abcd}\phi_{d}\nabla_{a}\tilde{Q}_{b}Q_{c}
\\
L_{A7}&=\varepsilon^{abcd}\phi_{d}\nabla_{a}\tilde{Q}_{b}\tilde{Q}_{c}
\\
L_{A8}&=\varepsilon^{abcd}\phi_{d}\nabla_{e}\tilde{Q}_{a}Q_{bc}{}^{e}
\\
L_{A9}&=\varepsilon^{abcd}\phi_{d}\nabla_{a}Q_{e}Q_{bc}{}^{e}
\\
L_{A10}&=\varepsilon^{abcd}\phi^{e}\nabla_{a}Q_{fbe}Q_{cd}{}^{f}
\\
L_{A11}&=\varepsilon^{abcd}\phi^{e}\nabla_{a}Q{}_{ebf}Q_{cd}{}^{f}
\\
L_{A12}&=\varepsilon^{abcd}\phi^{e}\nabla_{a}\tilde{Q}_{b}Q_{cde}
\end{aligned}
\qquad
\begin{aligned}[c]
L_{B1}&=\varepsilon^{abcd}\phi^{g}\phi_{g}\nabla_{e}Q_{ab}{}^{e}\nabla_{f}Q_{cd}{}^{f}
\\
L_{B2}&=\varepsilon^{abcd}\phi^{g}\phi_{g}\nabla_{f}Q_{ab}{}^{e}\nabla_{e}Q_{cd}{}^{f}
\\
L_{B3}&=\varepsilon^{abcd}\phi_{d}\phi^{g}\nabla_{e}Q_{ab}{}^{f}\nabla_{g}Q_{cf}{}^{e}
\\
L_{B4}&=\varepsilon^{abcd}\phi_{d}\phi_{g}\nabla^{e}Q_{ab}{}^{f}\nabla_{e}Q_{cf}{}^{g}
\\
L_{B5}&=\varepsilon^{abcd}\phi_{d}\phi_{g}\nabla^{e}Q_{ab}{}^{f}\nabla_{f}Q_{ce}{}^{g}
\\
L_{B6}&=\varepsilon^{abcd}\phi_{d}\phi_{g}\nabla^{g}Q_{ab}{}^{e}\nabla_{c}\tilde{Q}_{e}
\\
L_{B7}&=\varepsilon^{abcd}\phi_{d}\phi_{g}\nabla^{e}Q_{ab}{}^{g}\nabla_{c}\tilde{Q}_{e}
\\
L_{B8}&=\varepsilon^{abcd}\phi_{d}\phi^{g}\nabla_{e}Q_{ab}{}^{e}\nabla_{g}Q_{c}
\\
L_{B9}&=\varepsilon^{abcd}\phi_{d}\phi_{g}\nabla^{e}Q_{ab}{}^{g}\nabla_{e}Q_{c}
\\
L_{B10}&=\varepsilon^{abcd}\phi_{d}\phi_{g}\nabla^{g}Q_{ab}{}^{e}\nabla_{e}Q_{c}
\\
L_{B11}&=\varepsilon^{abcd}\phi_{d}\phi^{e}\nabla_{a}\tilde{Q}_{b}\nabla_{c}\tilde{Q}_{e}
\\
L_{B12}&=\varepsilon^{abcd}\phi_{g}\phi^{g}\nabla_{a}\tilde{Q}_{b}\nabla_{c}\tilde{Q}_{d}\,.
\end{aligned}
\end{equation}
Again, we observe that as $\phi_a=\dot{\phi}\delta^0_a$, the associated index must be temporal and that both $Q_{abc}$ and its (spatial) perturbation $q_{abc}$ cannot have $0$ in the last two indices. Moreover, as $\sqrt{-g}$ and $\phi_a$ are dependent only on time on the background FRW metric, we can liberally integrate by parts any derivative with spatial indices. Finally, as $\tilde{Q}^a$ vanishes on the background and its perturbation $\tilde{q}^a$ vanishes due to the transverse nature of the perturbation, we may omit any Lagrangian that can be expressed in terms of $\tilde{Q}^a$, while making use of the symmetries already established. Taking all these observations into account, we can greatly reduce the number of unique and non-vanishing Lagrangians to the following:
\begin{align}
\label{LAi}
\nn L_{A3} & =\varepsilon^{abcd}\phi_{d}\nabla_{a}Q_{fb}{}^{e}Q^{f}{}_{ce}\\
\nn L_{A4} & =\varepsilon^{abcd}\phi_{d}\nabla_{f}Q^{fe}{}_{a}Q_{bce}\\
L_{A11} & =\varepsilon^{abcd}\phi^{e}\nabla_{a}Q{}_{ebf}Q_{cd}{}^{f}.
\end{align}
For completeness, we include here the fourth-order terms that result from a straightforward generalisation of \eqref{L2deriv}. These are obtained by simply acting a $\nabla_g$ to each non-metricity tensor, resulting in the addition of two time derivatives to the Lagrangians:
\begin{equation}
L^{(4)}_{A3}=\varepsilon^{abcd}\phi_{c}\phi^{f}\nabla_g Q_{abe}\nabla^g Q_{fd}{}^{e},\qquad L^{(4)}_{A7}=\varepsilon^{abcd}\phi_{f}\phi^{f} \nabla_g Q_{abe}\nabla^g Q_{cd}{}^{e}\,.
\end{equation}
These, somewhat prosaic, higher-order generalisations of the non-vanishing second-order terms \eqref{L2deriv} would be suppressed by the necessary introduction of an energy scale. 

Further higher-derivative terms may only be obtained in this way i.e. a fifth-order term may be obtained through the generalisation $ L^{(5)}_{A3} =\varepsilon^{abcd}\phi_{d}\nabla_g\nabla_{a}Q_{fb}{}^{e}\nabla^gQ^{f}{}_{ce}$, etc., resulting in the addition of a further two time derivatives. Indeed, any odd-order in derivatives Lagrangian will be a generalisation of \eqref{LAi} and any even-order in derivatives Lagrangian will be a generalisation of \eqref{L2deriv}, see Appendix \ref{sec:nohigh}.

In accordance with the principle of effective field theory, these higher-order terms would be more suppressed order-by-order by an appropriate energy scale, which further motivates us to  restrict ourselves to Lagrangians containing (at most) two time derivatives\footnote{Note that (the symmetric teleparallel equivalent of) the Chern-Simons term is excluded from our consideration due to the quadratic restriction. Quadratic curvature invariants would correspond to quartic non-metricity invariants.}. As such, we are left with 3 unique Lagrangians that are of third order in derivatives \eqref{LAi}, while no four-derivative (or higher) theories remain. Thus, \eqref{LAi} combined with \eqref{L2deriv}, represents a complete characterisation of quadratic, parity-violating gravity in the STG formulation.
  \subsection{Propagation of GW}
  \label{propagation}

Recall that we have stipulated that the perturbation of the non-metricity tensor follows $Q_{ab}{ }^c\rightarrow \bar Q_{ab}{ }^c+\varepsilon q_{ab}{ }^c$, where $q_{ab}{}^{c}\equiv\delta Q_{ab}{}^{c}$. In perturbing the Lagrangians \eqref{LAi}, we must take care to perturb different forms of the non-metricity tensor correctly. To help in this regard, we state the following
\begin{equation}
\delta Q_{abc}=\partial_{a}h_{bc}\,,\qquad\delta Q^{f}{}_{ce}=\nabla^{f}h_{ce}\,,\qquad\delta Q^{fe}{}_{a}=\nabla^{f}h^{e}{}_{a}\,,\qquad\delta Q_{ab}{}^{c}=q_{ab}{}^{c}\,,
\end{equation}
where from \eqref{qabc1}, we have
\begin{equation}
q_{abc}=-2H\delta_{a}^{0}h_{bc}+\partial_{a}h_{bc}\,,\qquad q_{ab}{}^{d}=\partial_{a}h_{b}{}^{d}\,.
\end{equation}
These relations can be easily verified by standard means. Perturbing using the above identities, we find the remaining three-derivative Lagrangians to be of the form 
\begin{align}
L_{A3}	&=-\epsilon^{2}\varepsilon^{ijk}\dot{\phi}\partial_{a}\partial_{i}h_{j}{}^{l}\partial^{a}h_{kl}\,,
\nn\\
L_{A4}	&=-\epsilon^{2}\varepsilon^{ijk}\dot{\phi}\Box\partial_{i}h_{j}{}^{l}h_{kl}\,,
\nn\\
L_{A11}	&=-\epsilon^{2}\varepsilon^{ijk}\dot{\phi}\left(h_{k}{}^{l}\partial_{i}\ddot{h}_{jl}+\dot{h}_{k}{}^{l}\partial_{i}\dot{h}_{jl}\right)\,.
\end{align}
Unpacking these Lagrangians in terms of the spatial and temporal derivatives, we find that there are four unique terms contained within:
\begin{equation}
{\cal L}_{PV}^{(3)}\supset\varepsilon^{ijk}\partial^{2}h_{j}{}^{l}\partial_{i}h_{kl}\,,\quad\varepsilon^{ijk}\dot{h}_{j}{}^{l}\partial_{i}\dot{h}_{kl}\,,\quad\varepsilon^{ijk}\ddot{h}_{j}{}^{l}\partial_{i}h_{kl}\,,\quad\varepsilon^{ijk}{h}_{j}{}^{l}\partial_{i}\ddot{h}_{kl}\,.
\end{equation}
We may integrate the final two terms by parts to find that up to a function of $t$, the unique terms entering the parity-violating action will be
\begin{equation}
{\cal L}_{PV}^{(3)}\supset\varepsilon^{ijk}\partial^{2}h_{j}{}^{l}\partial_{i}h_{kl}\,,\quad\varepsilon^{ijk}\dot{h}_{j}{}^{l}\partial_{i}\dot{h}_{kl}\,,\quad\varepsilon^{ijk}\dot{h}_{j}{}^{l}\partial_{i}h_{kl}\,.
\end{equation}
Thus, the parity violating action to be analysed for the propagation of GW for a higher-derivative extension to General Relativity in the symmetric teleparallel geometry is given by
\begin{equation}
S^{(3)}=\frac{1}{2\kappa}\int dt\;d^{3}x\;a^{3}\left({\cal L}_{QGR}^{(2)}+\frac{\beta_{1}(t)}{a^{3}\Lambda}{\cal L}_{PV1}^{(3)}+\frac{\beta_{2}(t)}{a\Lambda}{\cal L}_{PV2}^{(3)}+\frac{\beta_{3}(t)}{a\Lambda}{\cal L}_{PV3}^{(3)}\right)\,,
\end{equation}
where $\beta_i(t)$ is a dimensionless function of time, $\Lambda$ is an energy scale and
\begin{equation}
{\cal L}_{PV1}^{(3)}\equiv\varepsilon^{ijk}\partial^{2}h_{j}{}^{l}\partial_{i}h_{kl}\,,\quad{\cal L}_{PV2}^{(3)}\equiv 2H\varepsilon^{ijk}\dot{h}_{j}{}^{l}\partial_{i}h_{kl}\,,\quad{\cal L}_{PV3}^{(3)}\equiv\varepsilon^{ijk}\dot{h}_{j}{}^{l}\partial_{i}\dot{h}_{kl}\,.
\end{equation}
Higher-derivative corrections require the introduction of a new energy scale in contrast to the previously studied case. 

\subsubsection*{Field Equations and conservation}
We find the contribution of the third order terms to the field equations to be given by 
\begin{equation}
\label{PV3eom}
\kappa\tau_{ij}^{(3)}=\frac{1}{2a^{2}}(h_{ij}^{\prime\prime}+2{\cal H}h_{ij}^{\prime}-\partial^{2}h_{ij})-\frac{1}{2a^{3}\Lambda}\varepsilon_{(ilk}\biggl[\left(-\beta_{1}\partial^{2}+\tilde{\beta}_{1}\right)\partial^{l}h^{k}{}_{j)}+\tilde{\beta}_{2}\bar{g}_{j)q}\partial^{l}h^{\prime kq}+\beta_{3}\bar{g}_{j)q}\partial^{l}h^{\prime\prime kq}\biggr]\,,
\end{equation}
where for presentation purposes, we have defined the functions
\begin{equation}
\tilde{\beta}_{1}=(\beta_{2}^{\prime}{\cal H}+\beta_{2}{\cal H}^{\prime})+3(\beta_{3}^{\prime}{\cal H}+\beta_{3}{\cal H}^{\prime})+\beta_{3}{\cal H}^{2}\quad \mbox{and}\quad\tilde{\beta}_{2}=\beta_{2}^{\prime}+3{\cal H}\beta_{2}\,,
\end{equation}
which are dependent only on conformal time $\eta$.
It is straightforward to verify that $\partial^i \tau^{(3)}_{ij} =\partial^j \tau^{(3)}_{ij}=0$ and that the trace vanishes, thus satisfying the conservation constraints derived in Section \ref{sec:cont}.

\subsubsection*{The wave equation}
\label{pgw}

Following the same procedure as in Section \ref{sec:GW2nd}, we find the higher derivative parity-violating extension to General Relativity to modify the propagation of GW like so
\begin{equation}
\left(1+\tilde{k}\lambda_{A}\beta_{3}\right)(h_{\vec{k}}^{A})^{\prime\prime}+\left(2+\tilde{k}\lambda_{A}\tilde{\beta}_{2}{\cal H}^{-1}\right){\cal H}(h_{\vec{k}}^{A})^{\prime}+\left(1+\tilde{k}\lambda_{A}(\beta_{1}+\tilde{\beta}_{1}k^{-2})\right)k^{2}(h_{\vec{k}}^{A})=0\,.
\end{equation}
We see here that all aspects of the propagation equation are modified by this higher derivative extension. We can combine the above with the second-derivative modification \eqref{GWs1} and express in terms of the general GW propagation equation \cite{Saltas:2014dha,Sawicki:2016klv,Nishizawa:2017nef} for a massless graviton like so
\begin{equation}
(h_{\vec{k}}^{A})^{\prime\prime}+(2+\nu_{A}){\cal H}(h_{\vec{k}}^{A})^{\prime}+(c_{T}^{A})^{2}k^{2}h_{\vec{k}}^{A}=0\,,
\end{equation}
where we have defined the effective Planck mass run rate or `friction' term as
\begin{equation}
\nu_{A}=\frac{\lambda_{A}\tilde{k}\left((\beta_{2}^{\prime}+3{\cal H}\beta_{2}){\cal H}^{-1}-2\beta_{3}\right)}{1+\beta_{3}\lambda_{A}\tilde{k}}\,,
\end{equation}
the GW propagation speed as\footnote{\textbf{Note:} that the dimension of the energy scale $\Lambda$ ensures correct dimensionality, i.e. Dim($\Lambda$)=Dim($k$)=Dim(${\cal H}$), where $a$, $\alpha$ and $\beta_i$ are dimensionless functions of proper time.}
\begin{equation}
c_{T}^{A}=\sqrt{\frac{1}{1+\tilde{k}\beta_{3}\lambda_{A}}\left[1+\lambda_{A}\left(\frac{4\alpha{\cal \tilde{H}}+\tilde{\gamma}^{(2)}}{\tilde{k}}+\beta_{1}\tilde{k}\right)\right]}\,,
\end{equation}
and $\gamma^{(2)}\equiv(\beta_{2}^{\prime}{\cal H}+\beta_{2}{\cal H}^{\prime})+3(\beta_{3}^{\prime}{\cal H}+\beta_{3}{\cal H}^{\prime})+\beta_{3}{\cal H}^{2}$, where the superscript $^{(2)}$ indicates the order of derivatives. The above parameters have been made dimensionless through the introduction of an energy scale $\Lambda$, i.e. $\tilde k\equiv k/ (a\Lambda)$, $\tilde{\cal H}\equiv {\cal H}/(a \Lambda)$ and $\tilde \gamma^{(2)}\equiv \gamma^{(2)}/(a\Lambda)^2$.

To return to General Relativity, the effective Planck mass run rate $\nu_A\rightarrow 0$ and the GW propagation speed $c_{T}\rightarrow 1$. At the latter limit, gravitational waves propagate at the speed of light and deviation from this is tightly constrained by recent observations from LIGO \cite{Abbott:2016blz}.

\subsubsection*{Propagation speed}
As ${\cal H}^2 \lll k^2 \lll 1$, we expand the propagation speed in series around $\tilde {\cal H}$ 
\begin{equation}
(c_{T}^{A})^{2}=\frac{1+\beta_{1}\lambda_{A}\tilde{k}}{1+\beta_{3}\lambda_{A}\tilde{k}}+\frac{4\alpha\lambda_{A}}{(1+\beta_{3}\lambda_{A}\tilde{k})\tilde{k}}\tilde{{\cal H}}+{\cal O}(\tilde{{\cal H}}^{2})\,,
\end{equation}
where we have noted that $\gamma^{(2)}\sim {\cal O}(\tilde{{\cal H}}^2)$. We now expand in series around  $\tilde{k}$ to obtain
\begin{equation}
\label{cTexpand}
c^A_{T}=1+\frac{1}{2}(\beta_{1}-\beta_{3})\lambda_{A}\tilde{k}+\frac{2\alpha\lambda_{A}}{k}{\cal \tilde{H}}-\alpha(\beta_{1}+\beta_{3})\lambda_{A}^{2}{\cal \tilde{H}}+{\cal O}^{2}(\tilde{k},\tilde{{\cal H})}\,.
\end{equation}
We see now that the dominant term
\begin{equation}
c_{T}^A=1+\frac{1}{2}\left(\beta_{1}-\beta_{3}\right)\lambda_A\tilde{k}+{\cal O}(\tilde{k}^{2})\,,
\end{equation}
is of precisely the same form as the propagation speed found in parity-violating Riemannian theories \cite{Nishizawa:2018srh}. By noting that $\rvert\lambda^A\rvert=1$, we may express the above as
\begin{equation}
\rvert-c^A_{T}+1\rvert=\frac{1}{2}\rvert\beta_{1}-\beta_{3}\rvert\tilde{k}\,,
\end{equation}
in order to compare with the propagation speed observed from the coincident detections GW170817/GRB170817A. The LIGO  experiment tightly constrains the propagation speed of GW within the bounds $-7\times10^{-16}<-c^A_{T}+1<3\times10^{-15}$, \cite{Abbott:2016blz,TheLIGOScientific:2017qsa,Monitor:2017mdv} and, as a result we find that parity-violating gravity in the STG geometry is constrained by~\footnote{Here we have taken the weaker constraint that results from assuming $-c_{T}+1>0$.} 
\begin{equation}
\rvert\beta_{1}-\beta_{3}\rvert\tilde{k}<6\times10^{-15}\,.
\end{equation}
Imposing the frequency $k/a\sim 100$Hz for consistency with the LIGO constraint we find
\begin{equation}
\Lambda^{-1}\rvert\beta_{1}-\beta_{3}\rvert<0.0912\text{ eV}^{-1}\qquad \mbox{i.e.}\qquad \Lambda>10.967\rvert\beta_{1}-\beta_{3}\rvert\text{ eV}\,.
\end{equation}
Thus, parity-violating corrections suppressed by an energy scale less than 10eV are ruled out. On the one hand, we cannot yet rule out corrections at the electroweak scale (let alone, of course the Planck scale), which could be considered theoretically most plausible. On the other hand, we can completely exclude the possibility that such corrections are related to dark energy (or dark matter) in such a way that they would appear at the same far infra-red regime. 

\subsubsection*{Two-derivative Parity-Violating Gravity}
From \eqref{cTexpand}, we observe that the two-derivative, parity-violating term $\alpha$ does not feature in the dominant term. In order to analyse this modification, we return to the two-derivative parity-violating gravity described in Section \ref{sec:GW2nd}, by setting $\beta_i=0$. This gives a vanishing run rate $\nu_A=0$ (as in GR) and a propagation speed of
\begin{equation}
c^A_{T}=\sqrt{1+\lambda_A\frac{4\alpha{\cal \tilde{H}}}{\tilde{k}}}=1+\frac{2\alpha\lambda_A\tilde{{\cal H}}}{\tilde{k}}+{\cal O}(\tilde{k},\tilde{{\cal H}}^{2})\,,
\end{equation}
from \eqref{cTexpand}, resulting in the bound
\begin{equation}
\frac{2\rvert\alpha\rvert\rvert\tilde{{\cal H}}\rvert}{\tilde{k}}<3\times10^{-15}\,.
\end{equation}
By inserting the latest value for the Hubble constant $H_{0}=70.3_{-5.0}^{+5.3}\text{km}\text{s}^{-1}\text{Mpc}^{-1}=1.43228\cdot10^{-33} \text{eV}$, we find an expression which constrains the dimensionless parameter $\alpha$ like so
\begin{equation}
\rvert\alpha\rvert< 10^{5}\,.
\end{equation}
As already deduced in the previous section, the lowest order parity-violating cannot be ruled out at cosmological scales by the GW data. This is because the effect of the parity violation is proportional to the current expansion rate of the universe. 

\section{Conclusion}
\label{sec:conc}

We considered parity-violating corrections to the symmetric teleparallel equivalent of General Relativity, and constrained their magnitude using the multi-messenger GW data. In symmetric teleparallel geometry, there is more freedom
to construct extensions of gravity theory without using higher derivatives. While all the metric extensions can be rewritten in the symmetric teleparallel geometry, the latter features more general possibilities that cannot be incorporated in the purely metric geometry. In particular, there is one quadratic, first-derivative extension described by the action\footnote{As shown in \cite{Iosifidis:2018zwo}, this parity-violating invariant happens to be projectively symmetric and is thus also covariant
under two definitions of a conformal transformation.}
\be \label{theory}
S_{QGR}= \frac{M^2}{2}\int d^4 x \sqrt{-g}\left( L_{QGR} + f\varepsilon^{abcd}Q_{abe}Q_{cd}{}^e\right)\,.
\ee
The form of $L_{QGR}$ was given in (\ref{qgrscalar}) and the function $f$ (which could be just a constant) depends on the theory. We found that the magnitude of $f$, which controls the proportion of the parity-violating corrections, is not efficiently constrained by the present GW data. If the special signature of the above correction was detected in the future data, that would mean that GW with left-handed polarisation reach the detectors
at a slightly different time than the GW with right-handed polarisation. This does not, in principle, require multi-messenger astronomy.

We also studied more general quadratic parity-violating corrections, attempting to parameterise the most general leading-order parity-violation according to the principle of effective field theory (which is simply that higher-than
quadratic-order corrections should be more suppressed). We found that there are three distinct terms that are relevant, and they are third order in derivatives and thus appear at the next-to-leading order. The energy scale associated to these terms is constrained by the present data, which imposes the lower bound of some dozen electronvolts. 
This lower bound is of the same order of magnitude as what one finds for the leading order parity-violating corrections in the metric geometry. This is in stark contrast to the case of the lowest order modification in (\ref{theory}), which does not have an equivalent in the purely metric gravity theory.

~
	\appendix
	\section{Variation of the affine connection:}
	\label{sec:Lie}
	
	Take the Lie derivative of the affine connection $\Gamma$ with respect to a vector $X$
	\begin{equation}
	{\cal L}_{X}\Gamma^{a}{}_{cb}=\partial_{c}\partial_{b}X^{a}+\Gamma^{a}{}_{cd}X^{d}{}_{,b}+\Gamma^{a}{}_{db}X^{d}{}_{,c}-\Gamma^{d}{}_{cb}X_{d}{}^{,a}+\Gamma^{a}{}_{cb,d}X^{d}\,.
	\end{equation}
	By noting the definition of the Riemann tensor
	\begin{equation}
	R^{a}{}_{bdc}=\partial_{d}\Gamma^{a}{}_{cb}-\partial_{c}\Gamma^{a}{}_{db}+\Gamma^{e}{}_{cb}\Gamma^{a}{}_{de}-\Gamma^{e}{}_{db}\Gamma^{a}{}_{ce}\,,
	\end{equation}
	the torsion tensor $T^{a}{}_{cb}=\Gamma^{a}{}_{[cb]}$ and following a number of straightforward manipulations, we arrive at the identity
	\begin{equation}
	{\cal L}_{X}\Gamma^{a}{}_{cb}=\nabla_{c}\nabla_{b}X^{a}+R^{a}{}_{bdc}X^{d}+\nabla_{c}\left(T^{a}{}_{db}X^{d}\right)\,.
	\end{equation}
	In the absence of curvature or torsion we find the Lie derivative of the affine connection to be simply
	\begin{equation}
	{\cal L}_{X}\Gamma^{a}{}_{cb}=\nabla_{c}\nabla_{b}X^{a}\,.
	\end{equation}
	As the change induced on a tensor field $\textbf{T}$ by an infinitesimal diffeomorphism is given by the Lie derivative along a vector field $X$~\footnote{For a discussion on the diffeomorphism invariance of STG, see \cite{us}.} , i.e. $\delta\textbf{T}=\epsilon {\cal L}_X \textbf{T}$, we arrive at a vital result in the calculation of the field equations for the connection, namely that the variation of the affine connection is given by
	\begin{equation}
	\label{delGamma}
	\delta \Gamma^d{ }_{ab}=\nabla_a\nabla_b \delta \xi^d\,,
	\end{equation}
	for some vector field $\xi^d$. By defining $\delta\xi^d\equiv \epsilon u^d$, we arrive at \eqref{delgammau} . Finally, we observe that in the absence of curvature and torsion we may commute the covariant derivatives, 
	in particular in the above equation.
	
	\section{Higher-derivative terms}
	\label{sec:nohigh}
	Let us consider 5-derivative, parity-violating gravity. The Lagrangian in such a theory must take the form
	\begin{equation}
	\varepsilon^{abcd}\phi\nabla\nabla Q\nabla Q\,,
	\end{equation}
	up to a factor of $\phi^2$, as before. In the class $\varepsilon^{abcd}\phi_d\nabla\nabla Q\nabla Q$, 	we can deduce a number of rules. Such Lagrangians will vanish if
		\begin{enumerate}
			\item ${abcd}$ appears twice in the last two indices of the non-metricity tensor $Q$, or
		\item ${abcd}$ appears twice in the $\nabla$'s or the first index of $Q$.
	\end{enumerate}
The last rule is a result of $\phi_d=\delta^t_d\phi_t$ which means that $abc$ are spatial, which allows for liberal integration by parts. This combined with the symmetry $\nabla_a Q_{bef}=\nabla_b Q_{aef}$ and the commutation of covariant derivatives, leaves only the terms
\begin{align}
\label{LAi5}
\nn L^{(5)}_{A3} & =\varepsilon^{abcd}\phi_{d}\nabla^g\nabla_{a}Q_{fb}{}^{e}\nabla_g Q^{f}{}_{ce}\,,\\
\nn L^{(5)}_{A4} & =\varepsilon^{abcd}\phi_{d}\nabla^g \nabla_{f}Q^{fe}{}_{a}\nabla_g Q_{bce}\,.
\end{align}
These are the straightforward generalisations of $L_{A3}$ and $L_{A4}$ given in \eqref{LAi} and discussed in Section \ref{sec:genhigh} and contain more than two time derivatives. These somewhat trivial higher-order terms are more suppressed by the appropriate energy scale and do not concern the present study.

For the class where $\varepsilon^{abcd}\phi^e\nabla\nabla Q\nabla Q$, $abcd$ can appear twice in the covariant derivatives or first index of $Q$, i.e. the Lagrangian must be of the form
\begin{equation}
\varepsilon^{abcd}\phi^{e}\nabla\nabla Q_{ac-}\nabla Q_{bd-}\,.
\end{equation}
Due to the traceless and transverse nature of the perturbations, any Lagrangian that can be rewritten to contain a $\tilde Q_a$ will vanish, leaving only
\begin{equation}
L_{A11}^{(5)}=\varepsilon^{abcd}\phi^{e}\nabla^{g}\nabla_{a}Q_{ebf}\nabla_{g}Q_{cd}{}^{f}\,,	
\end{equation}
which is the the straightforward generalisation of $L_{A11}$ found in \eqref{LAi} and contains more than two time derivatives.

Similarly, we may show that any 6th-order Lagrangians will be of the form
\begin{equation}
\varepsilon^{abcd}\phi\phi\nabla\nabla Q\nabla\nabla Q\,,
\end{equation}
and will be the straightforward generalisation of \eqref{L2deriv}. Indeed, any odd-order in derivatives Lagrangian will be a generalisation of \eqref{LAi} and any even-order in derivatives Lagrangian will be a generalisation of \eqref{L2deriv}.

\acknowledgements{The work was supported by the Estonian Research Council through the Personal Research Funding project PRG356 ``Gauge Gravity'' and by the European Regional Development Fund through the Center of Excellence TK133 ``The Dark Side of the Universe''.} 

	\bibliographystyle{unsrt}
	\bibliography{allcitations}
\end{document}